\documentclass[a4paper,11pt]{article}
\pdfoutput=1 

\usepackage{jcappub} 
\usepackage{bm}

\def\q {\bm{q}}
\def\x {\bm{x}}
\def\k {\bm{k}}

\def\d {{d}\,}
\newcommand{\fnl}{{f}_\text{NL}}

\newcommand{\HH}{\mathcal{H}}

\newcommand{\two}{^{\text{\tiny ({{2}})}}}
\newcommand{\one}{^{\text{\tiny ({{1}})}}}

\newcommand{\<}{\langle}
\renewcommand{\>}{\rangle}
\newcommand{\red}[1]{{#1}}
\newcommand{\apjl}{\rm{ApJ}}

\title{\boldmath The galaxy bias at second order in general relativity with Non-Gaussian initial conditions}
\author[a]{Obinna Umeh,}
\author[a]{Kazuya Koyama}

\affiliation[a]{Institute of Cosmology \& Gravitation, University of Portsmouth, Portsmouth PO1 3FX, United Kingdom}

\emailAdd{obinna.umeh@port.ac.uk}
\emailAdd{kazuya.koyama@port.ac.uk }

\abstract{We present a systematic study of galaxy bias in the presence of primordial non-Gaussianity in General Relativity (GR) at second order in perturbation theory. The non-linearity of the Poisson equation in GR and primordial non-Gaussianity are consistently included. We show that the inclusion of non-local primordial non-Gaussianity in addition to local non-Gaussianity is important to show the absence of the modulation of small scale clustering by the long-wavelength mode in the single field slow-roll inflation. We study the bispectrum of the relativistic galaxy density in several gauges and identify the effect of primordial non-Gaussianity and GR corrections.}

\begin{document}
\maketitle
\flushbottom

\section{Introduction}\label{sec:intro}
Our current understanding of the initial conditions of the Universe is mainly based on the measurements of temperature anisotropies of the Cosmic Microwave Background Radiation (CMBR) \cite{Ade:2015xua,Aghanim:2018eyx}. The CMBR analysis shows that the initial density field is almost Gaussian, however, there is still a large uncertainty in its determination of the amplitude of primordial non-Gaussianity \cite{Ade:2015ava}. Future Large Scale Structure (LSS) surveys  such as EUCLID~\cite{Amendola:2016saw}, LSST~\cite{Zhan:2017uwu} and SKA~\cite{Maartens:2015mra,Santos:2015gra}, as well as cross-correlations between these surveys,  promise to tighten the constraint on the amplitude of primordial non-Gaussianity to less than 1\%~\cite{Fonseca:2015laa,Alonso:2015uua}. 

The primordial local non-Gaussianity modulates \red{the formation of galaxies} and induces a scale dependence in the galaxy bias on \red{near horizon scales} \cite{Dalal:2007cu,Verde:2009hy,Assassi:2015fma}. The scale dependence provides a novel way to constrain primordial non-Gaussianity using LSS. Probing primordial non-Gaussianity with LSS requires an accurate modelling of the formation and evolution of tracers of LSS within General Relativity (GR)~\cite{Yoo:2012se,Camera:2014bwa,Castorina:2018zfk}. 

On very large scales, the effective field theory \red{approach provides a framework} to study \red{the large scale evolution of tracers of the underlying dark matter density field~\cite{Desjacques:2011jb,Assassi:2015fma,Desjacques:2016bnm}.  Within this framework for a vanishing pressure and anisotropic stress tensor, the  primordial non-Gaussianity generated by the standard single-field slow-roll inflation model~\cite{Maldacena:2002vr} does not leave any observable imprint on galaxy clustering
~\cite{Pajer:2013ana,dePutter:2015vga,Cabass:2016cgp}.} This implies that any detection of primordial non-Gaussianity in the squeezed limit will rule out single field slow-roll inflation model.

The study of \red{the effect of} non-Gaussian initial conditions on galaxy bias in GR has so far been limited to linear order~\cite{Baldauf:2011bh,Jeong:2011as,Bruni:2011ta,Challinor:2011bk}, and an extension to second order is limited to the Newtonian approximation of gravity~\cite{Assassi:2015fma,Desjacques:2016bnm}. In Ref.~\cite{Umeh:2019qyd} (Paper I), we identified GR corrections to the relativistic galaxy density with the Gaussian initial condition. In \red{current} paper, we include primordial non-Gaussianity and study the bispectrum of the relativistic galaxy density in several gauges to highlight the difference between primordial non-Gaussianity and GR corrections.  
 
We treat galaxies as biased tracers of the total mass distribution, which is dominated by dark matter~\cite{Kaiser:1984ApJ}. In GR, this implies that the \red{initial galaxy bias must be specified} in a gauge where both dark matter and galaxies are comoving,
i.e Comoving-Synchronous gauge (C-gauge)~\cite{Bartolo:2015qva}.
\red{The C-gauge defines a unique Lagrangian frame in GR~\cite{Ehlers:1993gf,Bertacca:2015mca}. The  C-gauge dark matter density we adopt here follows \cite{Matarrese:1997ay,Villa:2015ppa} but differs from the definition of comoving gauge given in \cite{Noh:2003yg,Yoo:2015uxa}, which allows a non-vanishing shift vector.}

\red{Another key ingredient in formulating galaxy bias in GR is the equivalence principle; the effect of perturbations with wavelengths greater than the galaxy formation scale, $R_g$, is simply to rescale small scale coordinates and it has no effect on the local dynamics~\cite{Pajer:2013ana,Dai:2015rda,Umeh:2019qyd}. Therefore, we define appropriate local coordinates where the long wavelength part of the initial curvature perturbation is absorbed as a part of the local coordinates \footnote{The details on how to construct local coordinates are given in \cite{Umeh:2019qyd} and it agrees with the Conformal Fermi Coordinates (CFC) introduced in \cite{Dai:2015rda}}. Finally, we adopt the effective theory formalism to decompose the dark matter density into long and short wavelength modes and smooth over the short modes to describe the evolution of the long-wavelength mode while hiding our ignorance of small scale physics such as the tidal effects in the bias parameters.} 

This paper is organised as follows. In section \ref{sec:modulation}, we introduce the Poisson equation in global coordinates in GR and show that the short wavelength mode of the matter density is modulated by the long-wavelength mode in local coordinates only with primordial non-Gaussianity beyond the standard single-field slow-roll inflation. In section \ref{sec:nonGaussiangalaxy}, we derive the galaxy bias model with a non-Gaussian initial condition in C-gauge and then show how to transform it to Eulerian frames in section \eqref{sec:galaxybiasEulerian}. We study the bispectrum of the relativistic galaxy density in section \ref{sec:Resultsanddission} and conclude in section \ref{sec:conc}.

\noindent
{\bf{Notations}}: 
%
We consider a universe which consists of dark matter and the cosmological constant only, i.e we ignore the effects of radiation and anisotropic stress tensor.  
\red{The perturbation theory expansion of any quantity $X$ is normalized as follows}:
$
X = \bar{X} + X\one + X\two/2
$
where $\bar{X} $ denotes the FLRW  background component. We decompose each perturbed quantity at order $n$ into two parts $X^{(n)}  = X^{(n)}_{\rm{N}} + X^{(n)}_{\rm{GR}} $, where $X^{(n)}_{\rm{N}} $ denotes the Newtonian approximation of $X^{(n)}$, while $X^{(n)}_{\rm{GR}} $ denotes the general relativistic corrections. The world-lines in C-gauge are labelled by the comoving coordinates ${\q}$.  We adopt the following values for the cosmological parameters~\cite{Ade:2015xua,Aghanim:2018eyx}: Hubble parameter, $h = 0.678$, baryon density parameter, $\Omega_b = 0.0485$,  dark matter density parameter, $\Omega_{\rm{cdm}} = 0.2595$, spectral index, $n_s = 0.9608$,  and the amplitude of the primordial perturbation, $A_s = 2.198 \times 10^{9}$. 

\section{Modulation of short mode of the matter density in General Relativity}\label{sec:modulation}

The initial conditions for the cosmological perturbation are set in terms of the gauge-invariant curvature perturbation, $\zeta_{{\rm ini}}$, on uniform density hypersurfaces. $\zeta_{{\rm ini}}$ is generated during inflation but remains constant (frozen) once it exits the horizon in a single field slow-roll inflation~\cite{Mukhanov:1990me,Wands:2000dp}. This allows
the initial conditions for matter-energy density to be fixed deep in the matter-dominated era in terms of $\zeta_{{\rm ini}}$. 
In the local non-Gaussianity model, the initial curvature perturbation up to second order in perturbation theory is given by~\cite{Komatsu:2002db}
\begin{eqnarray}
\zeta_{{\rm ini}}&=&\zeta_{{\rm ini}}\one+\frac{1}{2}\zeta_{{\rm ini}}\two=  \zeta_{{\rm ini}}\one+ \frac{5}{ 3}\fnl(\zeta_{{\rm ini}}\one)^2 + \cdots,
 \label{eq:initialconditions}
\end{eqnarray}
where $ \zeta_{{\rm ini}}\one$ denotes the linear part of the initial curvature perturbation, which is Gaussian.  $\fnl$ denotes the amplitude of local primordial non-Gaussianity. In general, there is also a non-local term that contributes to $\zeta_{{\rm ini}}\two$ which becomes important for the matter density via the Poisson equation. Including the non-local part of $\zeta_{{\rm ini}}$, we find \cite{Maldacena:2002vr,Acquaviva:2002ud}
\begin{eqnarray}
\zeta_{{\rm ini}}&=&\zeta_{{\rm ini}}\one+\frac{1}{2}\zeta_{{\rm ini}}\two=  \zeta_{{\rm ini}}\one+ \frac{5}{ 3}\fnl(\zeta_{{\rm ini}}\one)^2 + \frac{5}{ 3} \tilde{f}_{\rm{NL}} \partial^{-2}\left(\partial_{i}\zeta_{{\rm ini}} \partial^{i} \zeta_{{\rm ini}}\right)+ \cdots 
 \label{eq:initialconditions2}
\end{eqnarray}
\red{Similarly, $\tilde{f}_{\rm{NL}}$ denotes the amplitude of the non-local contribution to $\zeta_{{\rm ini}}$. In the single field slow-roll model, $\tilde{f}_{\rm{NL}}$ may be determined precisely in terms of the slow-roll parameters~\cite{Maldacena:2002vr,Acquaviva:2002ud}.} 
It is instructive to express $\zeta_{{\rm ini}}$ in terms of the initial gravitational potential $\Phi_{\rm{ini}}$
\begin{eqnarray}\label{eq:localform}
\Phi_{\rm ini}&=&\varphi_{{\rm ini}}\one+\fnl\,\left((\varphi\one_{{\rm ini}})^2 - \< \varphi^2_{{\rm ini}}\>\right) 
\\ \nonumber &&
+ \tilde{f}_{\rm{NL}}\left[ \partial^{-2}\left(\partial_{i}\varphi_{\rm ini} \one\partial^{i}\varphi_{\rm ini} \one\right) - \<\partial^{-2}\left(\partial_{i}\varphi_{\rm ini} \one\partial^{i}\varphi_{\rm ini} \one\right)\>\right] + \cdots
\end{eqnarray} 
where $\zeta_{\rm{ini}}\one = - {5} \varphi_{\rm{ini}}\one/{3}$ and $\varphi_{\rm{ini}}$ is the Gaussian limit of the initial gravitational potential.
Implementing the initial conditions and including the non-linear contributions from the perturbation of three-dimensional Ricci curvature~\cite{Bruni:2013qta,Gressel:2017htk} leads to a Poisson equation for the matter density field in a $\Lambda$CDM universe
\begin{eqnarray} \label{eq:Poissoneqnglobal}
    \nabla^2\Phi(\eta,{\q}) 
 & =&   \frac{3}{2} {\Omega_m \HH^2} \red{\delta_{{\rm{m N}}} }(\eta,{\q})
 \\ \nonumber 
 &+&\overbrace{\frac{10}{3} \bigg\{-
 \left[ \frac{1}{4}+ \frac{3}{5}\left(\fnl + \frac{1}{2}\tilde{f}_{\rm NL}\right) \right]
 \partial_i \varphi(\eta,{\q})\partial^i \varphi_{\rm{ini}}({\q})
    +\left( 1- \frac{3}{5}\fnl \right)  \varphi_{\rm{ini}}({\q}) \nabla^2 \varphi(\eta,{\q})\bigg\}}^{ \propto \,\delta_{{{\rm{GR,C}}}}}\,,
\end{eqnarray}
where $\HH\equiv \HH(\eta)$ is the conformal Hubble parameter, $\Omega_m\equiv \Omega_m(\eta)$ is the matter-energy density parameter and $\varphi_{\rm{ini}}({\q})$ is the initial time-independent potential. In the Gaussian limit of the Newtonian approximation, $\delta_{{{\rm{GR,C}}}}\rightarrow 0$; the Poisson equation becomes a linear equation at all orders in perturbation theory~\cite{Bernardeau:2001qr}. However, in GR, even in the Gaussian limit ($\fnl = \tilde{f}_{\rm{NL}}=0$), the Poisson equation is non-linear beyond the linear order in perturbation theory~\cite{Hidalgo:2013mba}. The GR non-linear correction is denoted as $\delta_{{{\rm{GR,C}}}}$. We make the following definitions to simplify the expressions; $\varphi({\q})\equiv \varphi(\eta, {\q})$, and $\delta_{\rm{m}}( {\q})\equiv \delta_{\rm{m}}(\eta, {\q})$.
In the limit  $\tilde{f}_{\rm NL} \rightarrow 0$, we recover the results presented in \cite{Hidalgo:2013mba,Villa:2015ppa}. The non-local primordial non-Gaussian term modifies the amplitude of $\partial_i \varphi({\q})\partial^i \varphi_{\rm{ini}}({\q})$ contribution in the Poisson equation.

With respect to the length scale of galaxy formation, $R_g$, we split the Poisson equation \eqref{eq:Poissoneqnglobal} into short and long wavelength modes, \red{the short wavelength mode component becomes}
\begin{eqnarray} \label{eq:Poissoneqnshort}
 \nabla^2\Phi_{s}({\q}) 
 &=&   \frac{3}{2} {\Omega_m \HH^2} \delta_{{\rm{Ns}}} ({\q})   
 \\   \nonumber && 
  +\frac{10}{3} \left( 1- \frac{3}{5}\fnl \right) \varphi_{{\rm{ini}}l}({\q}) \nabla^2 \varphi_{s}({\q})
 -\frac{20}{3} \left[ \frac{1}{4}+ \frac{3}{5}\left(\fnl + \frac{1}{2}\tilde{f}_{\rm NL}\right) \right]\partial_i \varphi_s({\q})\partial^i \varphi_{{\rm{ini}}l}({\q})
\,,
\end{eqnarray}
where we have neglected $\varphi_{{\rm{ini}}s}({\q})\nabla^2\varphi_{l}({\q}) ,\partial_i \varphi_s({\q})\partial^i \varphi_{{\rm{ini}}s}({\q}) $ and $\varphi_{{\rm{ini}}s}({\q}) \nabla^2 \varphi_{s}({\q})$ since they are sub-dominant and we focus on the coupling between the short and long modes.  
In this limit, it is clear that the short wavelength mode of the matter density is modulated by the long mode coming from two physical origins: primordial non-Gaussianity and non-linearity in GR. The latter contribution remains even in the limit $\fnl \rightarrow 0$ and $\tilde{f}_{\rm NL}  \rightarrow 0$.

\red{We showed in Paper I} that the contribution due to the non-linearity of GR to the matter density field can be removed by a large gauge transformation. This gauge transformation is related to the non-linearly realised symmetries, which give rise to the consistency relations~\cite{Hui:2018cag}. Similarly, the contribution of  $\fnl \varphi_{{\rm{ini}}l}({\q}) \nabla^2 \varphi_{s}({\q})$  to the short mode of matter density can be removed using the dilatation part of the large gauge transformation in \red{the single field slow-roll} inflation model~\cite{dePutter:2015vga}. It is still an open question as to whether all the long/short mode coupling given in equation \eqref{eq:Poissoneqnshort} can be removed by the residual gauge transformation in the single field slow-roll inflation. 
We shall explore this in detail by studying also the contribution from the gradient term in the presence of non-local primordial non-Gaussianity. 

The action for the curvature perturbation in the single field slow-roll inflation is invariant under the following transformation \cite{Goldberger:2013rsa}
\begin{eqnarray}\label{eq:etacoord}
\tilde{\eta}& =& \eta \,,
\\\label{eq:xcoordbody}
\tilde{q}^i &=&q^i (1- \frac{5}{3}\varphi^{\rm{ini}}_{0})  -\frac{5}{3} q^i {q}^j \partial_j \varphi^{\rm{ini}}_{0}  + \frac{5}{6} q_i q^i\partial^i \varphi^{\rm{ini}}_{0}\,,
\end{eqnarray}
where $ \varphi^{\rm{ini}}_0$ is the leading order term in the gradient expansion of $\varphi_{\rm{ini}}({\q})$ evaluated at the peak of the dark matter density field~\cite{Sheth:1999mn};
\begin{eqnarray}
 {\varphi_{{\rm{ini}}}}({\q}) &=& \varphi_{\rm{ini}}(\bm{0})+( \partial_{i}{\varphi}_{\rm{ini}})_{\bm{0}} q^{i} 
  + \frac{1}{2}(\partial_{i}\partial_{j}\varphi_{\rm{ini}} )_{\bm{0}} q^{i}q^{j}  + \mathcal{O}({\q})^{3} \,.
  \label{eq:Taylorseries}
\end{eqnarray} 
Here we have defined $\varphi_{0}^{\rm{ini}} \equiv \varphi_{\rm{ini}}(\bm{0})$, $ \partial_{i}{\varphi^{\rm{ini}}_0} \equiv\left( \partial_{i}{\varphi_{\rm{ini}}}\right)(\bm{0})$  and $\partial_{i}\partial_{j}{\varphi^{\rm{ini}}_0} \equiv \left(\partial_{i}\partial_{j}{\varphi_{\rm{ini}}}\right)(\bm{0})$. The coordinate transformation given in equation \eqref{eq:xcoordbody} is consistent with Conformal Fermi Coordinate (CFC) construction given in~\cite{Pajer:2013ana,Cabass:2016cgp}. \red{For more information on the explicit comparison between CFC construction and equations \eqref{eq:xcoordbody} \& \eqref{eq:etacoord}, see Appendix B of Paper I.}
The second term in equation \eqref{eq:xcoordbody} corresponds to the Dilatation (denoted as ${\rm{D}}$) while the last two terms correspond to the special conformal transformation (denoted as ${\rm{K}}$). Under these transformations, the short mode of the matter density field transforms in Fourier space as
\begin{eqnarray}
\delta_{\rm{D}}\delta_{\rm{s}}(\k)  &=& \frac{5}{3}\varphi^{\rm{ini}}_{0} \left[3 + {k^j} {\partial_{k_j}}\right]{\delta_{\rm{s}}(\k)}\,,
\\
\delta_{\rm{K}}\delta_{\rm{s}}(\k) &=& i \frac{5}{3}\partial^j \varphi^{\rm{ini}}_{0}\left[ 6 {\partial_{k_j} }+ 2{k^i}{{\partial_{k^i} \partial_{ k^j}} } -{k^j}{\partial^2_{k}} \right]\delta_{\rm{s}}(\k) \,.
\end{eqnarray}
Further simplification of these expressions shows that $\delta_{{\rm{s}}} ({\q})$ transforms as\footnote{For details on the simplification, see Paper I (\cite{Umeh:2019qyd}) }
\begin{eqnarray}\label{eq:transformation}
{\delta}_{\rm{s}} ( \tilde{\q})  &=&\delta_{\rm{s}} ({\q})+  \frac{5}{3} \left(2
+ \frac{d {\rm{log}} (k^3 \varphi_{s{\k}})}{d {\rm{log}} k} \right)\varphi^{\rm{ini}}_{0} \delta_{\rm{s}}\left({\q}\right) + \frac{5}{3}\left(1 
 - \frac{1}{2}\mathcal{Y} \right)\partial^j\nabla^{-2} \delta_{\rm{s}}( {\q})  \partial_j  \varphi^{\rm{ini}}_{0} \,.
 \label{eq:consistency}
\end{eqnarray}
The second term on the right-hand side corresponds to the response of the short mode density to the dilatation, while the last term results from the special conformal transformation.
The term $\mathcal{Y}$ is given by
\begin{eqnarray}
\mathcal{Y} =\frac{d^2\log (k^3 \varphi_{{s{\k}}})}{d(\log k)^2} +\left[\frac{d\log (k^3 \varphi_{{s{\k}}})}{d\log k} \right]^2+ \frac{d\log (k^3 \varphi_{{s{\k}}})}{d\log k} \,,
\end{eqnarray}
and it vanishes in the scale invariant limit~\cite{Umeh:2019qyd}. The non-linear terms in equation \eqref{eq:transformation} are generated purely from the coordinate transformation given in equation \eqref{eq:xcoordbody}. 

We shall now compare equation \eqref{eq:transformation} to equation \eqref{eq:Poissoneqnshort}. Evaluating equation \eqref{eq:Poissoneqnshort} at the peak of the matter density, the non-linear part of the matter density contrast becomes
\begin{eqnarray}\nonumber
  \frac{1}{2} \delta_{{{\rm{GR,C}}}(sl)}\two({\q})&= &  \frac{10}{3} 
  \left( 1- \frac{3}{5}\fnl \right)  \delta_{s}({\q})  \varphi^{{\rm{ini}}}_0 
  + \frac{20}{3} \bigg[
  \left( 1- \frac{d\log (k^3\varphi_{s{\bf{k}}})}{d\log k}\right) \left( 1- \frac{3}{5}\fnl \right)
    \nonumber\\ &&
- \left[ \frac{1}{4}+ \frac{3}{5}\left(\fnl + \frac{1}{2}\tilde{f}_{\rm NL}\right) \right] \bigg]
 \partial_i  \varphi^{{\rm{ini}}}_0  \partial^i \nabla^{-2} \delta_{s}( {\q}) \,,
\label{eq:perturbationtheory}
\end{eqnarray}
where we have made use of the Taylor expansion given in equation \eqref{eq:Taylorseries}
\begin{eqnarray}
\delta_{\rm{m}}({\q}) \varphi_{{\rm{ini}}l}({\q}) 
&\approx & \delta_{s}({\q}) \varphi^{{\rm{ini}}}_0 + \partial_{i}\varphi^{{\rm{ini}}}_0 \partial^{i} \nabla^{-2} \delta_{s}({\q})\left[ 1- \frac{d\log (k^3\varphi_{s{\bf{k}}})}{d\log k}\right].
\label{eq:phidelta}
\end{eqnarray}
We now show that (\ref{eq:perturbationtheory}) agrees with (\ref{eq:consistency}) in the single field slow-roll inflation. At the leading order in slow-roll parameters, the primordial non-Gaussian parameters are given by \cite{Maldacena:2002vr, Goldberger:2013rsa}
\begin{equation}
\fnl = - \frac{5}{6}\frac{d\log (k^3\varphi_{{\bf{k}}})}{d\log k}
= \frac{5}{12} (1- n_s) \ll 1, \quad  \tilde{f}_{\rm{NL}} = - {3} \fnl/{2}, 
\end{equation}
where $n_s$ is the spectrum tilt of the power specturm of the initial curvature perturbation. At the leading order in slow-roll parameters, 
$\mathcal{Y}$ is given by 
\begin{eqnarray}
\mathcal{Y} = \frac{d\log (k^3 \varphi_{{s{\k}}})}{d\log k} \,.
\end{eqnarray}
Then it is straightforward to show that (\ref{eq:perturbationtheory}) agrees with (\ref{eq:consistency}). This means that in the single field slow-roll inflation, the second order GR contribution is solely generated by the coordinate transformation. This shows also that the GR contribution can be removed by the coordinate transformation and it does not contribute to the formation of galaxies on small scales. On the other hand, in multi-field models, the primordial non-Gaussian contributions $\fnl$ and $ \tilde{f}_{\rm{NL}}$ cannot be removed by the local coordinate transformation and they will affect the galaxy bias. 
 
\section{Galaxy density in Lagrangian frame with relativistic corrections}\label{sec:nonGaussiangalaxy}
\red{We consider the limit where} the galaxy number density obeys a conservation equation\footnote{\red{For more discussion of the conservation of the galaxy number in GR context, see Paper I}}~{\cite{Desjacques:2016bnm,Umeh:2019qyd}}, in this limit, the galaxy density contrast at ${\tau}$ is given in terms of the galaxy density contrast at a formation time according to 
\begin{eqnarray}\label{eq:conservationgen}
 1+ \delta_{g}({{\q}})& =&
 \left[ 1 + \delta^{\text{L}}_{g}({\q})\right] \left(1 +  \delta_{\rm{m}}({{\q}})  \right) \,,
\end{eqnarray}
where $\delta^{\text{L}}_{g}({\q})=\delta_{g}({\tau}_{\rm{ini}},{\q})$ denotes the initial galaxy density contrast or the Lagrangian density. The question left to be answered is how to specify $ \delta^{\text{L}}_{g}$ in terms of the primordial curvature perturbation or the primordial gravitational potential $\varphi_{\rm{ini}}$. We have shown in section \ref{sec:modulation} that the effect of the homogeneous potential $\varphi^{\rm{ini}}_0$ and constant gradient $\partial_i \varphi^{\rm{ini}}_0$ can be removed by coordinate re-definition in the single field \red{slow-roll} inflation model. 
This implies that $\delta_{\rm{g}}^{\rm{L}} $ can only be expressed as a functional of terms with more than one spatial derivative of  $\varphi_{\rm{ini}}$; $\delta_{\rm{g}}^{\rm{L}} \sim \mathcal{F}[\partial_{i}\partial_{j}\varphi_{0}^{\rm{ini}},\partial_{i}\partial_{j}\partial_k\varphi_{0}^{\rm{ini}} + \cdots  ]$ in the single field \red{slow-roll} inflation. At the leading order, we construct observables from the irreducible decomposition of $\partial_{i}\partial_{j}\varphi^{\rm{ini}}_{0}$;
\begin{eqnarray}
\partial_{i}\partial_{j}\varphi^{\rm{ini}}_{0}  = \frac{1}{3}\nabla^2 \varphi_{0}^{\rm{ini}}\delta_{ij}^K  + s_{0ij} \,,
\label{eq:irrducibledecomp}
\end{eqnarray}
where $s_{0ij} = {D}_{ij}\varphi^{\rm{ini}}_{0}$ \red{with ${D}_{ij} = \partial_i \partial_j - \delta_{ij}^K\nabla^2/3$} is the initial tidal tensor~(i.e the electric part of the Weyl tensor \cite{Ellis2009}) and $\nabla^2 \varphi_{0}^{\rm{ini}}$ is related to the initial matter density via the Poisson equation. Therefore, $\delta^{\text{L}}_{g}$ can only be specified as a functional of $\delta_{{\rm{m}}}^{{\rm{ini}}}$ and $s^2_{\rm{ini}} =s_{0ij}s^{0ij} $(both indices of $s_{0ij}$ must be contracted because of the symmetry of the FLRW background~\cite{McDonald:2009dh})~\cite{Umeh:2019qyd}:
\begin{eqnarray}
\delta^{\text{L}}_{g}(\tilde{\q}) &\propto &\mathcal{F}[ \left(\delta_{{\rm{m}}}^{\rm{ini}}(\tilde{\q}),s^2_{\rm{ini}}(\tilde{\q})\right)]
\end{eqnarray}
However, if the inflationary model deviates from the standard single field slow-roll model,
a non-vanishing coupling between long and short wavelength modes is induced in the matter density field. In this case, we may include the dependence on $\varphi_{{\rm{ini}}}$ when we specify  $\delta^{\text{L}}_{g}(\tilde{\q})$ 
\begin{eqnarray}
\delta^{\text{L}}_{g}(\tilde{\q}) 
&\propto &\mathcal{F}[ \left(\left[\delta_{s}(\tilde{\q}) + \delta_{l}\one(\tilde{\q})\right],\left[\varphi_{{\rm{ini}},s}(\tilde{\q}) +\varphi_{{\rm{ini}},l}(\tilde{\q})\right],\left[s^2_s(\tilde{\q}) +s^2_l(\tilde{\q})\right]\right)]\,.
\end{eqnarray}
We have decompose each contribution into long and short wavelength modes.
We can now perform a series expansion in the long mode 
\begin{eqnarray}\label{eq:lagbiasexp}
\delta_{gl}^\text{L}(\tilde{\q})&=&b_{10}^\text{L}\delta_{l}\one(\tilde{\q})+b_{01}^\text{L}
\varphi_{{\rm{ini}},l}(\tilde{\q})
\\ \nonumber &&
+\frac{1}{2}\bigg[2b_{11}
^\text{L}\delta_{l}\one(\tilde{\q})\varphi_{{\rm{ini}},l}(\tilde{\q})+{ b_
{ 20} ^\text{L}}\left(\delta_{l}\one(\tilde{\q})\right)^2+{b_{02}^\text{L}}(\varphi_{{\rm{ini}},l}(\tilde{\q}))^2
+{ {b_{s} ^\text{L}}}s^2_l(\tilde{\q})\bigg]\,,
\end{eqnarray} 
\red{According to the effective field theory approach, the contribution of the short mode is hidden in the Lagrangian bias parameters. These bias parameters are coarse-grained representations of the physics of the short modes responsible for galaxy formation}.
We have adopted the two-index notation introduced in \cite{Tasinato:2013vna} for the Lagrangian bias parameters:
\begin{eqnarray}\label{eq:biasdef}
b_{ij}^{\text{L}} &\equiv&\left(\frac{D_{\rm{ini}}}{D}\right)^{i}  \left[  \frac{\partial^{i+j} \mathcal{F}\left[ \<\left(\delta_{s},\varphi_{{\rm{ini}},s},s^2_s)\>_{R_g}\right)\right]}{\partial \delta^i_l\partial \varphi^j_{{\rm{ini}},l} }\right]\bigg|_{\delta_l =\varphi_{{\rm{ini}},l} = 0}\,,
\\
 b_{s}^{\text{L}} & \equiv&\left(\frac{D_{\rm{ini}}}{D}\right)^2 \left[ \frac{\partial \mathcal{F}\left[ \<\left(\delta_{s},\varphi_{{\rm{ini}},s},s^2_s)\>_{R_g}\right)\right]}{\partial (s^2_l)}\right]\bigg|_{s^2_l = 0}\,.
\end{eqnarray}
The indices $i$ and $j$ correspond to the number of derivatives. \red{$D_{\rm{ini}}$ and $D$ are the growth factor at the initial and at a later time, respectively}. We did not use the two-index notation for the tidal bias since it appears only at second order. Substituting equation \eqref{eq:lagbiasexp} in equation \eqref{eq:conservationgen}
in local coordinates gives 
\begin{eqnarray}\nonumber \label{eq:galaxibias}
 \delta_{\!{\rm{gC}}l}( \tilde{\q})& =& \left[1 + b_{10}^{\rm{L}}\right]\delta_{\l}\one( \tilde{\q}) +  b_{01}^{\rm{L}}\varphi_{{\rm{ini}},l}(\tilde{\q})
\\ \nonumber &&
+ \frac{1}{2} \bigg[\delta\two_{{\rm {mN,C}} \l} (\tilde{\q}) +\delta\two_{{\rm{mGR,C}} \l}(\tilde{\q})+  \left[b_{20}^{\rm{L}} +2 b_{10}^{\rm{L}}\right]\left(\delta_{\l}\one(\tilde{\q})\right)^2 
\\  && \hspace{1cm}
+ 2\left[b_{01}^{\rm{L}} + b_{11}^{\rm{L}}\right]\delta_{\l}\one( \tilde{\q})\varphi_{{\rm{ini}},l}(\tilde{\q}) + b_{02}^{\rm{L}}\left(\varphi_{{\rm{ini}},l}(\tilde{\q})\right)^2
+  b_{s^2}^{\rm{L}}s^2_{\l}(\tilde{\q})
\bigg]\,,
\end{eqnarray}
where we have substituted for $\delta_{g}^{\rm{L}}$ in equation \eqref{eq:conservationgen} using equation \eqref{eq:lagbiasexp}. Here $\delta\two_{{\rm{mGR,C}} l}$ is the long-wavelength part of the GR correction to the matter density in C-gauge;
\begin{eqnarray}
 \delta_{{\rm{m\rm{GR,C}}l}}\two(\tilde{\q})&= &  \frac{20}{3} 
  \left( 1- \frac{3}{5}\fnl \right) \delta_{l}\one(\tilde{\q})  { \varphi_{{\rm ini},l}({\tilde{\q}})}
  \\ \nonumber &&
  - \frac{20}{3} \left[ \frac{1}{4}+ \frac{3}{5}\left(\fnl + \frac{1}{2}\tilde{f}_{\rm NL}\right) \right]
 \partial_i \nabla^{-2} \delta_{l}\one( \tilde{\q}) { \partial_i\varphi_{{\rm ini},l}({\tilde{\q}})}\,.
\label{eq:GRCorrections2}
\end{eqnarray}
We can obtain the galaxy density in global coordinates by performing a coordinate transformation of equation \eqref{eq:galaxibias}: $ \delta_{{\rm{gC}} l }({\q}) = \delta_{{\rm{gC}} l} (\tilde{\q}) - \partial_j \delta_{{\rm{gC}}l}\left({\q}\right)  \left(  \tilde{q}^j-{q}^j\right).
$
Applying the same coordinate re-mapping to the matter density $\delta_{l}(\tilde{\q}) $ and requiring that equation \eqref{eq:galaxibias} holds order by order, i.e $ \partial_j \delta_{{\rm{gC} }l } =\left[1 + b_{10}^{\rm{L}}(\tau)\right](\partial_{i}{\delta_{l}\one})+ b_{01}^{\rm{L}}(\tau)\partial_i\varphi_{{\rm ini},l}$, we obtain
\begin{eqnarray}\label{eq:Euleriangalaxydensity}
\delta_{{\rm{gC}}}({\q})&=& \left[1 + b_{10}^{\rm{L}}\right] \left[\delta_{\rm{m}}\one({\q}) + \frac{1}{2}\delta_{\rm{mN,C}}\two({\q}) \right] +  b_{01}^{\rm{L}}\varphi_{{\rm ini}} ({\q})
\\ \nonumber &&
+ \frac{1}{2}\bigg\{\left[ b_{20}^{\text{L}} + \frac{2}{3}b^{\text{L}} _{10}\left(1-\frac{{F}}{{D}^2}\right) \right](\delta_{\rm{m}}\one({\q}))^2
+ b_{02}^{\text{L}}(\varphi_{{\rm ini}}({\q}))^2 
\\ \nonumber &&
 + \delta\two_{{\rm{mGR,C}} }( \tilde{\q})+ 2\left[b_{01}^{\rm{L}} + b_{11}^{\rm{L}}\right] \delta_{\rm{m}}\one({\q}) \varphi_{{\rm ini}}({\q}) 
+ \left[  b_{s}^{\text{L}} -b_{10}^\text{L} \left(1-\frac{{F}}{{D}^2}\right)\right]s^{2}({\q})
\bigg\}\,,
\end{eqnarray}
where we omitted the subscript $l$ since all the terms are given by the long mode and set $\delta_{l}\one( {\q}) = \delta_{\rm{m}}\one({\q})$, which agrees exactly with the Newtonian expression at linear order. \red{To obtain equation \eqref{eq:Euleriangalaxydensity} in the present form, we made use of the second order perturbation theory expression for $\delta_{\rm{mN,C}}\two({\q})$
\begin{eqnarray}
\delta_{\rm{mN,C}}\two({\q})&= &
\frac{2}{3} \left(2+ \frac{F}{D^2}\right)  (\delta_{\rm{m}} \one({\q}))^2
+  \left(1-\frac{F}{D^2}\right) s^2({\q})
\end{eqnarray}
so that $\delta_{\rm{mN,C}}\two({\q})$ has the same form as its linear order counterpart. $F$ is the second-order growth factor, which obeys the second order differential equation
\begin{eqnarray}
 F'' +\mathcal{H}F'-\frac{3}{2} {\mathcal{H}^2\Omega_{\rm m}} F=\frac{3}{2}  {\mathcal{H}^2\Omega_{\rm m}} {D}^2 \,.
\end{eqnarray}
In the Einstein de-Sitter limit, it is related to the growth factor as $F = 3 D^2/7$.}
\section{Galaxy density in Eulerian frame with relativistic corrections}\label{sec:galaxybiasEulerian}
In GR, perturbations change under a general coordinate transformation
\begin{eqnarray}
 {x}^{\mu} \rightarrow x^{\mu} + Z^{\mu}\,, \qquad {\rm{with}} \qquad  Z^{\mu} = ( T,L^i).
 \label{eq:guagetransform}
\end{eqnarray}
Here $T$ stands for a temporal gauge transformation and $L^i$ corresponds to a spatial gauge transformation. We decompose $L^i =\partial^i L+ \beta^i $, where $\partial _i \beta^i =0$.  We consider two different gauge choices which correspond to the Eulerian frame; the Total matter gauge (T-gauge) and the Poisson gauge (P-gauge). 
The gauge transformation of the matter-energy density field from $C$-gauge to these gauges up to second order in perturbation theory is given by \cite{Bruni:1996im,Villa:2015ppa}. In the case of T-gauge, the scalar part of the gauge transformation at first order is given by \cite{Villa:2015ppa}
\begin{eqnarray}\label{eq:gaugegenCT1}
L\one({\x})&=&  \frac{2}{3} \frac{\varphi({\x})}{\mathcal{H}^2\Omega_{\rm m}}= \nabla^{-2}\delta_{\rm{m}}\one({\x}) \,, \qquad \qquad 
T = 0\,.\label{eq:gaugegenCT1B}
\end{eqnarray}
At the second order, the galaxy density and the primordial gravitational potential in C-gauge transform to the corresponding expression in T-gauge as \cite{Tellarini:2016sgp}
\begin{eqnarray}
\delta\one_{\rm{gC}}({\x}) &= &  \delta\one_{\rm{gT}}({\q})+{\partial^j\nabla^{-2} \delta\one_{\rm{m}}}({\x}) \partial_j \delta\one_{\rm{gC}}({\x})\,,
\\
\varphi_{{\rm ini}}({\x}) &= &  \varphi_{{\rm ini}}({\q})+{\partial^j\nabla^{-2} \delta\one_{\rm{m}}}({\x}) \partial_j\varphi_{{\rm ini}}({\x})
\,.
\end{eqnarray}
We can now rewrite the galaxy density in a more compact form by introducing  the Eulerian bias parameters
\begin{eqnarray}\label{eq:Tgaugegalaxybias}
\delta_{\text{gT}}({\x}) &=&b_{10}\left[\delta_{\rm{m}}\one({\x}) +\frac{1}{2}\delta_{\rm{mN,T}}\two({\x}) \right] +b_{01}\varphi_{{\rm ini}}({\x})+ \frac{1}{2}b_{02}\big[\varphi_{{\rm ini}}({\x})\big]^2
\\ \nonumber &&
+\frac{1}{2}\bigg[b_{20}\big[\delta\one_{{\rm m}}({\x})\big]^2 +b_{s^2}{s^2} ({\x}) +b_{11}\delta_{m}\one({\x})\varphi_{{\rm ini}}({\x})
+  b_{n^2}\partial^j\nabla^{-2}\delta\one_{\rm{m}}({\x}) \partial_j \varphi_{{\rm ini}} ({\x}) \bigg] \,,
\end{eqnarray}
where $\delta_{\rm{mN,T}}\two$ is the Newtonian approximation of the matter density contrast in T-gauge
\begin{eqnarray}\label{eq:TguageNewtonian}
\delta_{\rm{mN,T}}\two({\x})&= &
2\partial_j \nabla^{-2}\delta_{\rm{m}}\one({\x}) \partial_j \delta_{\rm{m}}\one({\x})+ \frac{2}{3} \left(2+ \frac{F}{D^2}\right)  (\delta_{\rm{m}} \one({\x}))^2
+  \left(1-\frac{F}{D^2}\right) s^2({\x}) \,,
\end{eqnarray}
and 
$\{b_{10}\,,  b_{01},b_{20},b_{11},b_{02}, b_{s^2},b_{n^2}\}$ are Eulerian bias parameters defined as a function of the Lagrangian bias parameters: 
\begin{eqnarray}\label{eq:Eb10}
b_{10}&=&1+b_{10}^\text{L}\,,\\
b_{20}&=&b_{20}^{\text{L}} + \frac{2}{3}\left(b_{10}-1\right)\left(1+ {{ F}\over { D}^2} \right)\,,\label{eq:Eb20}\\
 b_{s^2}&=& b_{s^{2}}^{\text{L}} -\left(b_{10} -1\right)\left(1+ {{F}\over { D}^2} \right)\,,\label{eq:bsq}\\
b_{01}&=& b_{01}^\text{L}\,,\\
 b_{11} &=& 2 \left[b_{11}^\text{L}  +  b_{01}^\text{L}\right] + \frac{20}{3} 
  \left( 1- \frac{3}{5}\fnl \right) \,, 
 \label{eq:dilatonbias} \\
 b_{02} &=&b_{02}^\text{L}\,,\\
  b_{n^2} &=&-2b_{01}^{\rm{L}} -\frac{20}{3} \left[ \frac{1}{4}+ \frac{3}{5}\left(\fnl + \frac{1}{2}\tilde{f}_{\rm NL}\right) \right]\,.
  \label{eq:vsqbias}
\end{eqnarray}
Equations \eqref{eq:Eb20}  and \eqref{eq:bsq} reduce to well-known expressions for the non-linear bias and tidal parameters in the Einstein de Sitter limit by setting ${ F} = 3 D^2/7$. 
We obtain the Newtonian result in the Eulerian frame by replacing equation \eqref{eq:dilatonbias} and equation \eqref{eq:vsqbias} by \cite{Baldauf:2010vn,Tellarini:2015faa}
\begin{eqnarray}
 b_{11}^{\rm{N}}&=& 2 \left(b_{11}^\text{L}  +  b_{01}^\text{L}\right) -4\fnl \,, \label{eq:Newtonianb11}
 \\
   b_{n^2}^{\rm{N}} &=&-2b_{01}^{\rm{L}} -  4\left(\fnl + \frac{1}{2}\tilde{f}_{\rm NL}\right) \,.
   \label{eq:Newtonianvsqbias}
\end{eqnarray}
In the Gaussian limit, this set of bias parameters vanishes $\{b_{01},  b_{11}, b_{02} \} \rightarrow 0$ and we are left with
\begin{eqnarray}
b_{10}&=&1+b_{10}^\text{L}\,,\\
b_{20}&=&\frac{8}{21}\left(b_{10}-1\right)+b_{20}^\text{L}\,,\\
 b_{s^2}&=&b_{s^{2}}^\text{L}-\frac{4}{7}\left(b_{10} -1\right)\,,\\
 \label{eq:Gaussianb11}
  b_{11}&=&\frac{20}{3}   \,,\\
   \label{eq:Gaussianbnsq}
  b_{n^2} &=& -\frac{5}{3} \,.
\end{eqnarray} 
These results agree with \cite{Umeh:2019qyd}.
For the single field \red{slow-roll} inflation models all the bias parameters that depend on the short mode component of $\varphi_{\rm{ini}}$  vanish and the galaxy density becomes
\begin{eqnarray}\label{eq:Tgaugegalaxybias2}
\delta_{\text{gT}}({\x}) &=&b_{10}\left[\delta_{\rm{m}}\one({\x}) +\frac{1}{2}\delta_{\rm{mN,T}}\two({\x}) \right] 
+\frac{1}{2}\bigg[b_{20}\big[\delta\one_{{\rm m}}({\x})\big]^2 +b_{s^2}{s^2} ({\x})  
\\ \nonumber &&+ \frac{20}{3} 
  \left( 1- \frac{3}{5}\fnl \right)\delta_{m}\one({\x})\varphi_{{\rm ini}}({\x})
- \frac{20}{3} \left[ \frac{1}{4}+ \frac{3}{5}\left(\fnl + \frac{1}{2}\tilde{f}_{\rm NL}\right) \right]\partial^j\nabla^{-2}\delta\one_{\rm{m}}({\x}) \partial_j \varphi_{{\rm ini}} ({\x})\bigg]  \,,
\end{eqnarray}
where the last line still carries information about the primordial non-Gaussianity and non-linearity of GR.  This correction arises due to distortions of the volume element determined by the  matter-energy density as the tracer evolve from the formation time to when it is opbserved~\cite{Umeh:2019qyd}. There is no modulation of small scale physics by the long mode and the bias parameters are not affected by primordial non-Gaussianity nor non-linearity of GR in the single field \red{slow-roll} inflation model.  In the case of unbiased tracers with $b_{10}=1, b_{20}=b_{s^2}=0$, this expression agrees with the dark matter density in T-gauge as expected.

Next, we consider the galaxy density in the Poisson gauge (P-gauge). The gauge transformation from C-gauge to P-gauge for the galaxy density at linear order is given by \cite{Bertacca:2014hwa,Jolicoeur:2017eyi}
\begin{eqnarray}
  \delta_{\rm{gP} }\one({\x}) 
&=& b_{10}\delta_{\rm{mNT}}\one({\x})  +b_{01}\varphi_{{\rm in}}({\x}) +3\HH v\one({\x})\,,
\label{eq:Poissonguagegalaxydenisty1}
\end{eqnarray}
where $v\one$ is the velocity potential. We neglect the contribution from the evolution bias since \red{we assumed the galaxy number conservation}. At the second order, the galaxy density in P-gauge is given by
\begin{eqnarray}\label{eq:Poissonguagegalaxydenisty2}
 \delta_{\rm{gP} }\two({\x})&=&b_{10}\delta_{\rm{mNT}}\two({\x}) + b_{02}\big[\varphi_{{\rm in}}({\x})\big]^2
+b_{20}\big[\delta_{{\rm m}}\one({\x})\big]^2 +b_{s^2}{s^2} ({\x}) +b_{11}\delta_{m}\one({\x})\varphi_{{\rm in}}({\x})
\\ \nonumber &&
+  b_{n^2}\partial^j\nabla^{-2}\delta\one_{\rm{m}}({\x}) \partial_j \varphi_{{\rm in}} ({\x}) 
+3\HH v\two({\x}) - {3}\left[  \HH'  - 4 \HH^2 \right] \big[v\one({\x})\big]^2 
- {6}\HH   v\one({\x}) \Phi\one({\x})
\\   \nonumber &&
+2v\one({\x})\bigg(3\HH \left(b_{10} \delta_m\one({\x}) + b_{01}\varphi_{{\rm in}}({\x}) \right)
-  {b_{10}'}\delta_m\one({\x}) -{{b_{01}'}}\varphi_{{\rm in}}({\x})
+b_{10} f \HH\delta_m\one({\x}) \bigg)\,,
\end{eqnarray}
where $\Phi$ is the Bardeen potential in P-gauge and we made use of the second order expression for the galaxy density given in \cite{Bertacca:2014hwa,Jolicoeur:2017eyi} and implemented the galaxy bias model given in equation \eqref{eq:Tgaugegalaxybias}. Note that we made use of the linear order Euler equation to substitute for $v'$,
\begin{eqnarray}
{v\one}^\prime({\x}) =- \HH v\one({\x}) - \Phi\one({\x}),  
\end{eqnarray}
and use equation \eqref{eq:Poissonguagegalaxydenisty1} to obtain 
$
{\delta\one_{\rm{gT}}}'({\x})= \big[{b_{10}'}+b_{10} f \HH\big]\delta_m\one({\x}) +  {{b_{01}'}}\varphi_{{\rm in}}({\x})\,
$
and we introduced the growth rate defined as $D^{\prime} = D f \HH$.   

\section{Linear and non-linear scale dependent bias}\label{sec:scaledependentbias}
We show how the scale dependent bias emerges at first and second order in perturbation theory with primordial non-Gaussianity. First, we expand the galaxy density in Fourier space;
\begin{eqnarray}
\delta_{g_{\rm{X}}}\one({\k}) &=& \int \frac{\d^{3}k_1}{(2\pi)^{3}}{{\mathcal{K}}\one_{\rm{X} } ({\k}_{1})} \delta\one_{{\rm{m}}}({\k}_{1})(2\pi)^3\delta^{D}({\k}_1-{\k})\,,
\\ \label{eq:Fourierdelta2}
\delta_{g_{\rm{X}}}\two({\k}_3)&=& \int\frac{d^3 k_{1}}{(2\pi)^{3}}\int {d^3 k_{2}}{\mathcal{K}}\two_{\rm{X} } {({\k}_{1}, {\k}_{2},{\k}_3)}\,\delta\one_{{\rm{m}}}({\k}_{1})\delta\one_{{\rm{m}}}({\k}_{2})
\delta^{D}({\k}_{1}+{\k}_{2}-{\k}_{3})
\\ \nonumber && 
\qquad - \<\delta_{g\rm{X}}\>\delta^{D}({\k}_3)\,,
\end{eqnarray}
where $X=P, T$ stands for a different gauge and ${\mathcal{K}}\one_{\rm{X} }$ is a gauge dependent kernel. The specific form of $\mathcal{K}_{X}$ is obtained by expanding the galaxy density in Fourier space. In equation \eqref{eq:Fourierdelta2}, we subtracted off the ensemble average of $\delta_{g_{\rm{P}}}\two$ to ensure that $\<\delta_{g_{\rm{P}}}\> = 0$~\cite{Gil-Marin:2014baa}.
For the galaxy density in T-gauge,  the kernel is obtained by expanding equation \eqref{eq:Tgaugegalaxybias} in Fourier space
\begin{eqnarray}
{\mathcal{K}}\one_{\rm{T} } ({\k})&=&b_{\rm{T}}\one({\k})\,,\\
{\mathcal{K}}\two_{\rm{T} } ({\k}_{1}, {\k}_{2},{\k}_3)&=&b\two_{\rm{T}}({\k}_1,{\k}_2) + b_{10} F_{2\rm{N}}({\k}_1,{\k}_2)+b_{s^2}S_2({\k}_1,{\k}_2)
+ b_{n^2}\mathcal{N}_2({\k}_1,{\k}_2) \,.
\label{eq:KernelT2}
\end{eqnarray}
where $b_{\rm{T}}\one({\k})$ is a scale dependent linear bias parameter. It is a linear combination of $b_{10}\delta_{\rm{m}}\one({\x})$  and $b_{01}\varphi_{\rm in}({\x})$ obtained from the first line of equation \eqref{eq:Tgaugegalaxybias}
\begin{eqnarray}\label{eq:scaledependentb1}
 b_{\rm{T}}\one({\k})&\equiv &b_{10} + \frac{b_{01}
}{\alpha(z,{k})}\,.
\end{eqnarray}
Assuming a simple halo bias model from where galaxy bias parameters may easily be calculated~\cite{Gonzalez-Perez:2017mvf}, equation \eqref{eq:scaledependentb1} reduces to the well-known linear scale dependent bias parameter with $b_{01}\propto2 \fnl \left(b_{10}-1\right)$ \cite{Verde:2009hy,WandsandSlosar}. Following \cite{Baldauf:2010vn}, we express $\varphi_{\rm ini}({\x})$ in terms of the evolved matter density
\begin{eqnarray}
\varphi_{\rm{ini}}(k) =\frac{\delta_m(z,k)}{\alpha(z,k)}\,,
\end{eqnarray}
where we introduced an auxiliary function
\begin{eqnarray}\label{eq:auxillaryfunction}
 \alpha(z,k) \equiv-\frac{2}{3} \frac{k^2T(k)}{\Omega_m \HH^2}\frac{g}{g_{\rm{ini}}}=-\frac{10}{9} \frac{k^2T(k)}{\Omega_m \HH^2}\left(1 + \frac{2}{3}\frac{f}{\Omega_m}\right)^{-1} \,.
\end{eqnarray}
Using $a g = D$, $a \Omega_m \HH^2 = \Omega_{m0} \HH^2_{0}$ and $g(0)/g_{\rm{ini}}\sim 1.3$, we recover exactly the relation given in \cite{Baldauf:2010vn}.  Note that our $\varphi_{\rm in}$ is related to the the Bardeen potential $\Phi_{\rm in}$ according to $\Phi_{\rm in} = - \varphi_{\rm in}$~\cite{Verde:2009hy,Baldauf:2010vn,Villa:2015ppa}. From equation \eqref{eq:localform}, it implies that our $\fnl$ has an opposite sign compared with the $\fnl$ defined with respect to $\Phi_{\rm in}$.
Similarly, we obtain the scale dependent non-linear bias parameter 
\begin{eqnarray}
 b_{\rm{T}}\two({\k}_1,{\k}_2)&\equiv&  b_{20}  + \frac{b_{11}}{2}\left(\frac{\alpha(z,{k}_1) +\alpha(z,{k}_2)}{\alpha(z,{k}_2)\alpha(z,{k}_1)}\right) 
+\frac{b_{02}}{\alpha(z,{k}_1)\alpha(z,{k}_2)}\,.
\end{eqnarray}
This is a combination of the following terms $b_{20}[\delta_{{\rm m}}({\x})]^2 $, $ b_{11}\delta_{m}({\x})\varphi_{\rm in}({\x})$  and $b_{02}[\varphi_{\rm in}({\x})]^2$ . $ F_{2\rm{N}}({\k}_1,{\k}_2)$ and $S_2({\k}_1,{\k}_2)$ are the usual second order Newtonian matter density and tidal tensor kernels
\begin{eqnarray}
 F_{ 2N}({\k}_1,{\k}_2) &= &   \frac{10}{7} +\left(\frac{k_1}{k_2} +\frac{k_2}{k_1}\right)\frac{{\k}_1\cdot{\k}_2}{k_1k_2} +\frac{4}{7} \frac{({\k}_1\cdot {\k}_2)^2}{k_1^2k_2^2}\,,
 \\
 S_2({\k}_1,{\k}_2)& =& \frac{\left({
 \k}_1\cdot {\k}_2\right)^2}{\left(k_1 k_2\right)^2} - \frac{1}{3}\,,
\end{eqnarray}
where we have made an Einstein de Sitter assumption in $F_{ 2N}$.
Furthermore, $\mathcal{N}_2$ is given by
\begin{eqnarray}
 \mathcal{N}_2({\k}_1,{\k}_2)&=&\frac{1}{2}\left[\frac{{\k}_1\cdot {\k}_2}{k_1^2\alpha(z,{k}_2)}  + \frac{{\k}_1\cdot {\k}_2}{k_2^2\alpha(z,{k}_1)} \right]\,,
\end{eqnarray}
This term represents the effect of volume distortions due to a displacement of the galaxy position in the Eulerian frame. In the Newtonian limit, only the value of two bias parameters change
\begin{eqnarray}
b\two_{\rm{T}}({\k}_1,{\k}_2) &\rightarrow& b\two_{\rm{N}}({\k}_1,{\k}_2)= b_{20}  + \frac{b_{11}^{\rm{N}}}{2}\left(\frac{\alpha(z,{k}_1) +\alpha(z,{k}_2)}{\alpha(z,{k}_2)\alpha(z,{k}_1)}\right) 
+\frac{b_{02}}{\alpha(z,{k}_1)\alpha(z,{k}_2)}\,,
\\
 b_{n^2}^{\rm{T}} &\rightarrow &  b_{n^2}^{\rm{N}}\,,
\end{eqnarray}
where $b_{11}^{\rm{N}}$ and $b_{n^2}^{\rm{N}}$ are given in equation \eqref{eq:Newtonianb11} and \eqref{eq:Newtonianvsqbias}, respectively. In Poisson gauge, the kernels are obtained from equations \eqref{eq:Poissonguagegalaxydenisty1} and \eqref{eq:Poissonguagegalaxydenisty2}
\begin{eqnarray}
{\mathcal{K}}\one_{\rm{P} } ({\k})  &=&{\mathcal{K}}\one_{\rm{T} } ({\k})   + 3f\frac{\HH^2}{k^2} \,,
\\
{\mathcal{K}}\two_{\rm{P} } ({\k}_{1}, {\k}_{2},{\k}_3)
&=& {\mathcal{K}}\two_{\rm{T} } ({\k}_{1}, {\k}_{2},{\k}_3)+ 3f\frac{\HH^2}{k_3^2}
G_{2}({\k}_1,{\k}_2,{\k}_3)\,,
\end{eqnarray}
where $ v\one({k}) = \HH f \delta_{m}\one({\k})/k^2$ and, at second order, we introduced $G_{2}$, which is the Fourier space kernel for a collection of the peculiar velocity and its correlation with the galaxy density and gravitational potential
\begin{eqnarray}
G_{2}({\k}_1,{\k}_2,{\k}_3)=G_{2{\rm{N}}}({\k}_1,{\k}_2,{\k}_3) +G_{2\rm{GR}}({\k}_1,{\k}_2,{\k}_3) +G_{ 2\rm{Q}}{\k}_1,{\k}_2,{\k}_3) \,.
\end{eqnarray}
Here $G_{2{\rm{N}}}$ and $G_{2\rm{GR}}$ are the Newtonian and the GR correction to the Newtonian kernels of the intrinsic peculiar velocity contribution at second order, respectively and ${{G_{ 2\rm{Q}}}}$ is a collection of terms quadratic in first order terms
\begin{eqnarray}
G_{2\rm{N}}({\k}_1,{\k}_2,{\k}_3)
 &=& \frac{6}{7} + \left(\frac{k_1}{k_2}+ \frac{k_2}{k_1}\right) \frac{{\k}_1\cdot {\k}_2}{k_1k_2}
 + \frac{8}{7}\frac{({\k}_1\cdot{\k}_2)^2}{k_1^2 k_2^2} \,,
 \\ \label{eq:GGR2}
G_{2\rm{GR}}({\k}_1,{\k}_2,{\k}_3)&=&\frac{3}{2} {\Omega_{m}\HH^{2}}\bigg[\frac{k_3^2}{k_{1}^{2}k_{2}^{2}}\left[-3 + \frac{6}{5} \fnl \left(1 + \frac{2}{3}\frac{f}{\Omega_m}\right) \right]
\\ \nonumber &&
+ \frac{6}{5}\tilde{f}_{\rm{NL}}\left(1 + \frac{2}{3}\frac{f}{\Omega_m}\right)\frac{{\k}_{1}\cdot {\k}_{2}}{k_{1}^2k_{2}^2}
+\frac{2}{k_3^2}   E_{2}({\k}_1,{\k}_2)\bigg]\,,
  \\
G_{ 2\rm{Q}}({\k}_1,{\k}_2,{\k}_3) &=& \frac{2k_3^2}{k_{1}^{2}k_{2}^{2}}\bigg[\mathcal{Y}_2(k_1, k_2)+\frac{3}{2} {\Omega_{m}\HH^{2}}\bigg[\frac{ f}{\Omega_m}\left(1+\frac{1}{2}\Omega_m\right)
 +1\bigg]
 \bigg]\,.
\end{eqnarray}
Here we used $ \HH' = \HH^2 \left(1 - {3} \Omega_m/{2}\right)$, which is valid for a cosmological constant and matter dominated universe, and introduced the following definitions for clarity
\begin{eqnarray}
E_{2}({\k}_1,{\k}_2)&=& 3+2\frac{{\k}_{1}\cdot {\k}_{2}}{k_{1}k_{2}}\bigg(\frac{k_{1}}{k_{2}} + \frac{k_{2}}{k_{1}}\bigg) + \frac{({\k}_{1}\cdot {\k}_{2})^{2}}{k_{1}^{2}k_{2}^{2}}\,,
\\
\mathcal{Y}_2(k_1, k_2)&=&   \frac{1}{2}\left(b_{\rm{T}}\one({k}_2) k_1^2  +b_{\rm{T}}\one({k}_1) k_2^2 \right) + \frac{(1+z)}{6}\left(\frac{d {b_{\rm{T}}\one}({k}_2)}{d z} k_1^2  +\frac{d {b_{\rm{T}}\one}({k}_1)}{d z} k_2^2 \right)
\\ \nonumber &&
 + \frac{1}{6}b_{10} f\left(k_1^2 + k_2^2\right) \,,
\end{eqnarray}
Note that in P-gauge, in addition to the presence of $\fnl$ in the scale dependent linear and nonlinear bias parameters, i.e $b\two_{\rm{T}}$ and $b\one_{\rm{T}}$, the effect of the primordial non-Gaussianity is also contained in $G_{2\rm{GR}}$ and $\mathcal{Y}_2$. This is coming from the temporal component of the gauge transformation vector, in P-gauge, it corresponds to the peculiar velocity potential.

\section{Bispectrum}\label{sec:Resultsanddission}

\subsection{Galaxy bias from halo model}
There are 6 unknown Lagrangian bias parameters to be determined from observations or simulations. We can reduce the number of unknown Lagrangian bias parameters to 3 by invoking a halo model, i.e we assume that galaxies are resident in collapsed haloes of a certain mass range $d M$.
The number density of haloes of mass $M$ is given by \cite{Sheth:1999mn} 
\begin{equation}\label{eq:universalmassfct}
n(M)=\nu f_{\nu}(\nu)\frac{\bar \rho_{\rm{m}}}{M^2}\frac{\d \ln{\nu}}{\d \ln{M}}\,, \qquad \nu=\left(\frac{\delta_\text{c}}{\sigma_\text{nG}}\right)^2,
\end{equation}
where $\bar{\rho}_{\rm{m}}$ is the background matter density, $\nu f_{\nu}(\nu)$ is the multiplicity function~\cite{Sheth:1999mn}, $\nu$ denotes the peak height, and $\sigma_{\rm{nG}}$ is the variance in the matter density field smoothed on a Lagrangian mass scale $M$. For simplicity, we neglect the non-Gaussian correction to the variance in the matter density field $\sigma_{\rm{G}}\approx \sigma_{\rm{nG}}$ \cite{Baldauf:2010vn,Tellarini:2016sgp}.  Assuming a Sheth-Tormen (ST) model for $ \nu f_{\nu}(\nu)$ \cite{Sheth:2001dp}, it is straight forward to obtain the following Lagrangian halo bias parameters~\cite{Baldauf:2010vn}: $b_{h10}^{\rm{L}} \propto \partial n /({n} \partial \delta_l)$ and $b_{h20}^{\rm{L}} \propto \partial^2 n /({n} \partial \delta_l^2)$. The corresponding galaxy bias parameters are obtained by averaging over the halo bias parameters~\cite{Yankelevich:2018uaz}
\begin{equation}\label{eq:lagbias}
b_{ij}^{\rm{L}}=  \frac{\int_{M_{-}}^{M_{+}}  b^{\rm{L}}_{\!h ij}(M) \<N_g|M\>n(M) \d M } { \int_{M_{-}}^{M_+} \<N_g|M\> n(M)\d M}\,,
\end{equation}
where $ b^{\rm{L}}_{\!h ij}$ denotes the halo bias parameters, $\langle N_{\rm g}| M \rangle $ denotes the number of galaxies contained within a single dark matter halo of mass $M$,  $M_{-}$ and $M_{+}$ are the lower and upper mass limits of haloes that can host a particular type of galaxy. To calculate  $\langle N_{\rm g}| M \rangle $,  we consider a Euclid-type H-$\alpha$ galaxy survey and follow the modelling described in \cite{Yankelevich:2018uaz} which is based on the analysis given in \cite{Gonzalez-Perez:2017mvf}. Using equations \eqref{eq:Eb10} and \eqref{eq:Eb20} we find that the best fit `fundamental bias parameters' are given by
\begin{eqnarray}\label{eq:fitb10}
 b_{10} \left( z\right)&=& 0.9 + 0.4 z \, ,
 \\ \label{eq:fitb20}
 b_{20}\left(z\right)&=& -0.704172 -0.207993 z +0.183023 z^{2}-0.00771288 z^3 \, .
\end{eqnarray}
We neglect the initial tidal bias parameter and in this limit $b_s^2$ is easily obtained from equation \eqref{eq:fitb10} $b_s^2 = -4(b_{10} -1)/7$. Similarly, other bias parameters can easily be expressed in terms of equations \eqref{eq:fitb10} and \eqref{eq:fitb20};
\begin{eqnarray}
b_{01}&=&2 \fnl\delta_c \left(b_{10}-1\right)\,,\\
b_{02}&=&4\fnl^2 \delta_c\left[\delta_c b_{20}-2\left(\frac{4}{21}\delta_c+1\right)\left(b_{10}-1\right)\right]\,,
\\
b_{11}^\text{L}  &+&  b_{01}^\text{L}=2 \fnl \left[\delta_c b_{20}+\left(\frac{13}{21}\delta_c-1\right)\left(b_{10}-1\right)\right]\,,
\end{eqnarray}
where we followed the steps outlined in \cite{Baldauf:2010vn,Tellarini:2016sgp}. Note that these set of bias parameters vanish in the limit $\fnl \rightarrow 0$.

\subsection{Galaxy bispectrum}
General covariance requires that the descriptions of any observable quantity to be independent of coordinate systems. In cosmological perturbation theory, there is an additional requirement of gauge invariance associated with the mapping between the background spacetime and the physical spacetime. 
To construct observables such as the galaxy number count at non-linear order, one has to choose a coordinate system appropriate for an observer (i.e. past-lightcone) and compute the observables accordingly~\cite{Koyama:2018ttg}. In this paper, we will not discuss the observed galaxy bispectrum. Rather we compute the galaxy bispectrum $B_{{\rm{gX}}}$ of the relativistic galaxy density on the hypersurface of a constant time in a given gauge including the GR corrections and compare it to the Newtonian approximation. This is a gauge dependent quantity but it plays the central role when computing the observed galaxy bispectrum. The galaxy bispectrum in $X$ gauge is given by
\begin{eqnarray}
B_{{\rm{gX}} }({\k}_{1},  {\k}_{2},  {\k}_{3}) = \mathcal{K}_{X}\one({k}_{1}) \mathcal{K}_{X}\one({k}_{2}) \mathcal{K}_{X}\two({k}_{1},  {k}_{2},{k}_{3}) P_m(k_{1})P_m(k_{2}) +   \text{2 perms.} \,,
\end{eqnarray}
where $ P_m(k)\sim \<\delta_{m}\one({\k}) \delta_{m}\one({\k}') \>$ is the matter power spectrum, $\text{2 perms.}$ denotes two cyclic permutations and $\mathcal{K}_{X}$ is the Fourier space kernel for the galaxy density perturbations given in section \ref{sec:scaledependentbias}. 
For short wavelength modes that enter the horizon in the radiation dominated era, the analytic solutions for the second order perturbation cannot be used. For these modes, we need to use a numerical code such as a second order Einstein-Boltzmann code \texttt{SONG} \footnote{https://github.com/coccoinomane/song} or an improved analytic solution \cite{Tram:2016cpy}. This is particularly important for the squeezed configuration while the equilateral configuration is less affected by the radiation effects.  

\begin{figure}[h]
\centering 
\includegraphics[width=155mm,height=80mm] {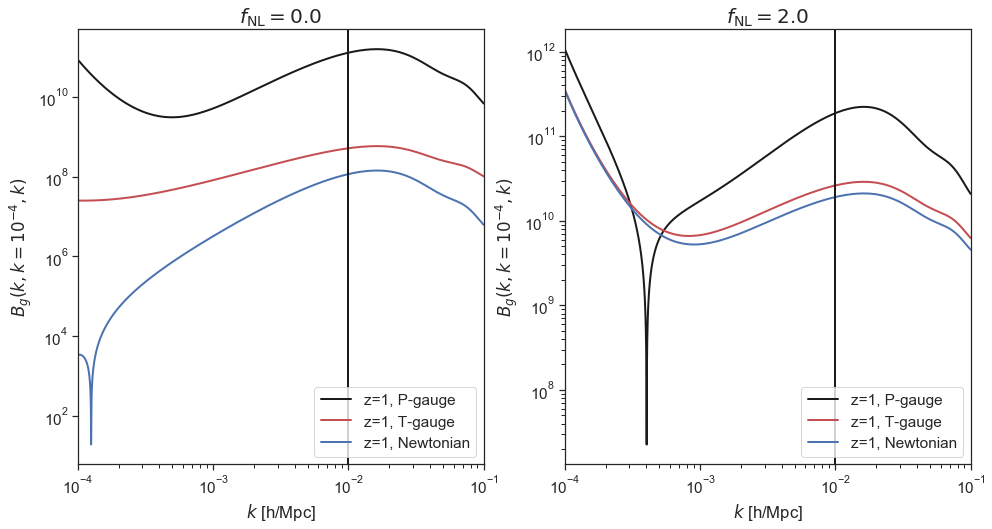}
\caption{\label{fig:sqezzedshape} The left panel shows the galaxy bispectrum in P-gauge, T-gauge and the Newtonian limit in Gaussian limit. Similarly in the right panel, we show the galaxy bispectrum for a non-Gaussian initial conditions with the $\fnl =2.0$ and  $\tilde{f}_{\rm{NL}}=0.0$. For both panels we fixed $k_2 = 0.0001[h\rm{Mpc}^{-1}]$ and vary $k_1=k_3=k$. Note that our result is valid only for $k<k_{\rm eq} \sim 0.01$ due to the fact that we ignore radiation effects.}
\end{figure}
We show in figure \ref{fig:sqezzedshape} the squeezed shape of the galaxy bispectrum in P-gauge and T-gauge. We compare the bispectrum in these two relativistic gauges to the galaxy bispectrum in the Newtonian limit. {{In the Gaussian limit, the bispectrum is negative except int the Newtonian limit at small $k$. This feature was first noticed in the Newtonian treatment of the halo bispectrum in \cite{Baldauf:2012hs}. This is due to the contribution of the non-linear bias parameter $b_{20}$; tracers with the negative $b_{20}$ leads to negative galaxy bispectrum. In the Newtonian limit, the bispectrum goes through zero at small $k$ and becomes positive when the tidal bias contribution is important. This feature is absent in the GR bispectrum in both gauges, because GR corrections induce an effective bias parameter $b_{11}={20}/{3}$ (see equation\eqref{eq:Gaussianb11}), which makes $b_{\rm{T}}\two$ more negative. }}
The contribution from  $b_{n^2} = -{5}/{3}$ vanishes in the squeezed limit for an equal time correlation. The bispectrum in P-gauge contains additional GR corrections from the temporal gauge transformation.  For a non-Gaussian initial condition (the right panel of figure \ref{fig:sqezzedshape}), the bispectrum in T-gauge and the Newtonian are similar once the primordial non-Gaussian contribution dominates over the GR correction. Again the bispectrum in P-gauge is significantly different due to additional GR {{and primordial non-Gaussian corrections in the temporal gauge transformation.}} 

	\begin{figure}[h]
\centering 
\includegraphics[width=155mm,height=80mm] {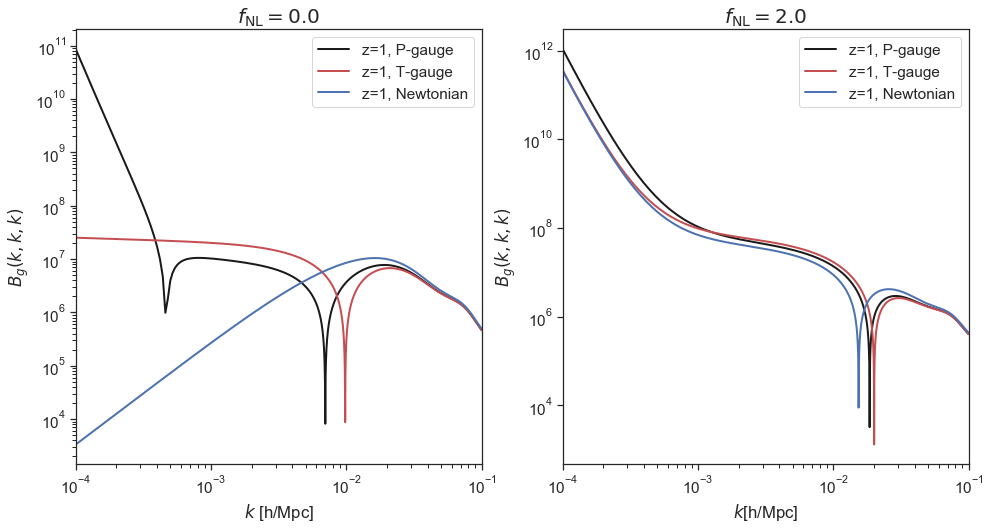}
\caption{\label{fig:eq} The left and right panels show the galaxy bispectrum in P-gauge, T-gauge and the Newtonian limit for the equilateral shape for the Gaussian and non-Gaussian initial conditions, respectively. Again we set  $\tilde{f}_{\rm{NL}}=0.0$.}
\end{figure}
In figure \ref{fig:eq}, we show the galaxy bispectrum in equilateral configurations. 
{{
For the equilateral shape in the Gaussian initial limit, the bispectrum in the Newtonian approximation is positive and decreases as $k \rightarrow 0$. }} However, the bispectrum in both relativistic gauges deviates strongly from the Newtonian prediction for $k < k_{\rm{eq}}$. This is due to the GR effects in the galaxy density, which modify {{both}} $b_{11}$ and $b_{n^2}$ bias parameters as discussed above. In particular, {{ $b_{11}  = 20/3$ causes that the galaxy bispectrum to become negative at small $k$ since $\alpha(z,k) \propto -k^2$. In P-gauge, it changes sign again due to the positive contribution from the temporal gauge transformation terms.}} The Newtonian approximation of the galaxy bispectrum also becomes negative with $\fnl \neq 0$ because $b_{11}^{N} =-4\fnl\left(b_{10} -1\right)$ is negative for $b_{10} >1$. The difference between the bispectrum in relativistic gauges and the Newtonian limit is small once the primordial non-Gaussianity effect becomes dominant over GR corrections{{ for $\fnl \geq 2$}}.

\section{Conclusion}\label{sec:conc}
We studied the galaxy bias in the presence of primordial non-Gaussianity in GR at second order in perturbation theory. Earlier studies at second order focused on the Newtonian approximation \cite{Baldauf:2010vn,Assassi:2015fma,Desjacques:2016bnm,Tellarini:2015faa} and an extension to GR was limited to the Gaussian limit~\cite{Umeh:2019qyd}. We used the local coordinates introduced in~\cite{Umeh:2019qyd}, which are equivalent to the conformal Fermi coordinates developed in \cite{Dai:2015rda,Dai:2015jaa}. \red{The exact map between the conformal Fermi coordinates and the local coordinates used here was given in~\cite{Umeh:2019qyd}}. 

We introduced a parametrisation of the initial curvature perturbation that includes local and non-local primordial non-Gaussianity contributions and used it to derive the full GR expression for the matter density and the peculiar velocity at second order in perturbation theory. 
We showed that the non-linear corrections to the Poisson equation in GR can be removed by local coordinate transformations in the single field slow-roll inflation model. In local coordinates, there is no modulation of small-scale clustering by the long-wavelength model. On the other hand, in multi-field inflation models, the primordial non-Gaussianity introduces the modulation of small scale clustering by the long mode. 

We then derived the {{local Lagrangian galaxy bias model}} in GR at the second order in perturbation theory in various gauges for a generic non-Gaussian initial condition. We found that the non-linearity of GR and primordial non-Gaussianity affect the galaxy density on large scales from distortions of the volume element by the long-wavelength mode. Also, primordial non-Gaussianity affects galaxy bias due to the modulation of small scale clustering by the long mode. In both squeezed and equilateral configuration, the galaxy bispectrum in the Poisson and total matter gauges differ substantially from the Newtonian approximation on ultra-large scales. The galaxy bias model that we derived in this paper is an essential building block for calculating the observed bispectrum of the galaxy number count and/or the HI brightness temperature~\cite{Umeh:2016nuh,Penin:2017xyf,
Umeh:2015gza,Umeh:2016thy,Jolicoeur:2017eyi,Bertacca:2017dzm,
Jolicoeur:2018blf,Koyama:2018ttg}.

\acknowledgments
We thank Marco Bruni, Robert Crittenden, Roy Maartens and  David Wands for useful discussions. Some of the tensor  algebraic computations here were done with the tensor algebra software xPand \cite{Pitrou:2013hga}.
OU and KK  are supported by the UK STFC grant ST/N000668/1 and ST/K0090X/1.  KK is also supported by the European Research Council under the European Union's Horizon 2020 programme (grant agreement No.646702 ``CosTesGrav").

\providecommand{\href}[2]{#2}\begingroup\raggedright\endgroup
\end{document}